%% file: 00.main.tex
\documentclass[sigconf]{acmart}

\usepackage{multirow}
\usepackage{color, colortbl}
\definecolor{Gray}{gray}{0.9}
\definecolor{LightGray}{gray}{0.6}

\hypersetup{colorlinks, citecolor={black}, filecolor=black, linkcolor=black, urlcolor=black}

\newcommand{\CC}[1]{\cellcolor{gray!#1}}

\usepackage{acmart-taps}

\AtBeginDocument{%
  \providecommand\BibTeX{{%
    \normalfont B\kern-0.5em{\scshape i\kern-0.25em b}\kern-0.8em\TeX}}}

\copyrightyear{2020}
\acmYear{2020}
\setcopyright{acmlicensed}\acmConference[MUM 2020]{19th International Conference on Mobile and Ubiquitous Multimedia}{November 22--25, 2020}{Essen, Germany}
\acmBooktitle{19th International Conference on Mobile and Ubiquitous Multimedia (MUM 2020), November 22--25, 2020, Essen, Germany}
\acmPrice{15.00}
\acmDOI{10.1145/3428361.3428463}
\acmISBN{978-1-4503-8870-2/20/11}

\begin{document}

\title{Alone or With Others? Understanding Eating Episodes of College Students with Mobile Sensing}

\author{Lakmal Meegahapola}
\affiliation{
  \institution{Idiap Research Institute and EPFL}
  \country{Switzerland}
}
\email{lmeegahapola@idiap.ch}

\author{Salvador Ruiz-Correa}
\affiliation{
  \institution{IPICYT}
  \country{Mexico}
}
\email{salvador.ruiz@ipicyt.edu.mx}

\author{Daniel Gatica-Perez}
\affiliation{
  \institution{Idiap Research Institute and EPFL}
  \country{Switzerland}
}
\email{gatica@idiap.ch}

\renewcommand{\shortauthors}{Meegahapola et al.}

\begin{abstract}
Understanding food consumption patterns and contexts using mobile sensing is fundamental to build mobile health applications that require minimal user interaction to generate mobile food diaries. Many available mobile food diaries, both commercial and in research, heavily rely on self-reports, and this dependency limits the long term adoption of these apps by people. The social context of eating (alone, with friends, with family, with a partner, etc.) is an important self-reported feature that influences aspects such as food type, psychological state while eating, and the amount of food, according to prior research in nutrition and behavioral sciences. In this work, we use two datasets regarding the everyday eating behavior of college students in two countries, namely Switzerland (N$_{ch}$=122) and Mexico (N$_{mx}$=84), to examine the relation between the social context of eating and passive sensing data from wearables and smartphones. Moreover, we design a classification task, namely inferring \textit{eating-alone vs. eating-with-others} episodes using passive sensing data and time of eating, obtaining accuracies between 77\% and 81\%. We believe that this is a first step towards understanding more complex social contexts related to food consumption using mobile sensing. 


\end{abstract}



\maketitle

\input{01.introduction}
\input{02.datasets}
\input{03.methodology}

\input{05.conclusion}

\newpage 

\bibliographystyle{ACM-Reference-Format}
\bibliography{citations}

\end{document}

%% file: 01.introduction.tex
\section{Introduction}\label{sec:introduction}


{Food and Nutrition} has risen as the second most common category of apps used by mHealth app users according to recent reports \cite{Mobius2019}. Most of these apps have already incorporated mobile food diaries/journals to provide basic temporal insights to users. Even though many food diary-based studies have been carried out in the past with encouraging results for food consumption related interventions \cite{Zepeda2008, Weiqing2019}, passive smartphone sensing has just begun to be widely used, comparatively speaking, in conjunction with food diaries, to understand contextual aspects that affect food consumption \cite{Biel2018, Seto2016}. 

Mobile food diaries use two main techniques to capture data \cite{Santani2018, Biel2018, Seto2016}. They are: (a) Passive Sensing - using embedded sensors in smartphones and wearables (accelerometer, gyroscope, location, etc.) and events generated from the phone (app usage, screen-on time, and battery charging events), models can unobtrusively generate behavioral and contextual insights; and (b) Self-Reports - capture details regarding daily behavior and context related to eating. According to prior literature in mobile sensing \cite{Biel2018, Seto2016} where eating is considered as a holistic event with interconnected dimensions \cite{Bisogni2007, Tulu2017}, the social context of eating is an important variable that is self-reported, as it is a factor that relates to many aspects regarding food consumption episodes such as location, time, psychological state while eating, physical conditions, and food amount.


Studies have found that eating in highly social contexts (partying, celebrations, gatherings, etc.) can influence the amount of food consumed, which might lead to overeating in the short term \cite{Bongers2013, Bruch1964, Patel2001} and to eating disorders in the long term \cite{Sawyer2000, Chua2004}. Further, concepts such as \textit{social facilitation} and \textit{impression management} emphasize how the presence of one or more people when eating can lead to overeating and undereating, respectively \cite{Herman2003, Herman2005, Hetherington2006}. Moreover, studies have examined the effects of \textit{eating-alone} and \textit{eating-with-others} as fundamental aspects regarding eating behavior \cite{Hetherington2006, Yates2017}. Hence, understanding the social context of eating has been outlined as an important component of food consumption research \cite{Haarman2020, Dunbar2017, Cruwys2015, Carroll2013, Tulu2017, Gemming2015}. Furthermore, automatically inferring attributes related to social context would enable mobile food diaries to send context-aware notifications \cite{Kandappu2020} and to support interventions \cite{Payne2015}, and also to help users adhere to healthy eating practices \cite{Gemming2015, Higgs2016}. In this study, similar to prior mHealth sensing studies with food diaries \cite{Biel2018, Seto2016, Meegahapola2020}, we consider eating to be a {holistic} event, and use a binary categorization for the social context of eating -- \textit{eating-alone} vs. \textit{eating-with-others} as a construct to understand food consumption behavior of college students in two countries. Hence, this paper has two contributions:

\noindent \textbf{Contribution 1:} We conducted a data analysis of wearable and mobile sensing datasets collected from 206 college students of two countries (Switzerland and Mexico). We highlight behavioral differences in the two countries, and show that factors such as time, physical activity levels, and location around eating episodes, all of which are passively sensed via wearables or smartphones, are features that help discriminate the social context of eating.

\noindent \textbf{Contribution 2:} We define and evaluate a classification task related to inferring the social context of eating (eating-alone vs. eating-with-others) using time of eating and passive sensing (wearable and smartphone) data related to physical activity levels and eating contexts, with accuracies in the range 77\%-81\%. These accuracies increase further when combined with other contextual features in both datasets. Hence, these initial results suggest that self-reported social context of eating can be inferred using sensing data with a above-baseline accuracy, showing the potential of using passive sensing together with mobile food diaries towards understanding more complex social contexts of eating. 

%% file: 02.datasets.tex
\section{Datasets and Pre-Processing}\label{sec:datasets}
The feature groups we used are temporal (\textbf{T}), contextual (\textbf{C}), and activity (\textbf{A}). Further, data sources are denoted by the by sub indices denoting self-reports 
 (\textit{$_{sr}$}), fitbit (\textit{$_{fb}$}), smartphone (\textit{$_{sp}$}), and other passive sensing modalities (\textit{$_{ps}$}). 

\textbf{Switzerland Dataset (CH-Dataset):}\label{subsec:ch_dataset}
We used a mobile sensing dataset called \emph{Bites'n'Bits} collected in our group’s previous work \cite{Biel2018, GaticaPerez2019}. It contains smartphone sensor data, self-reported data, and activity data of fitbit wearables from 122 students of a Swiss university. The smartphone application allowed users to self-report details regarding eating events in-situ (\textbf{T}: time of eating; \textbf{C$_{sr}$}: social context of eating, food type, concurrent activities, etc.). Further, their activity levels were captured using a fitbit wearable (\textbf{A$_{fb}$}: step count, physical activity level), and activity level features were derived using the minutes spent on each of the four levels: sedentary, lightly active, fairly active, and very active. Moreover, passive sensing data regarding the context such as location were captured (\textbf{C$_{ps}$}). In the final dataset, there are 4448 reports (3414 meals, 1034 snacks). All the participants in the study were between 18-26 in age, with a mean age of 20.5 years, and there were 65\% men and 35\% women.

\textbf{Mexico Dataset (MX-Dataset):}\label{ eating reportssubsec:mx_dataset}
We collected a dataset from 84 university student in San Luis Potosi, Mexico using the same approach described in \cite{Biel2018}. The dataset had self-reported features similar to CH-Dataset (\textbf{C$_{sr}$} and \textbf{T} feature groups), and instead of the FitBit wearable (which we left out for cost reasons), activity levels of students were captured using the phone accelerometer (\textbf{A$_{sp}$}: x, y, and z axis values of the accelerometer). Further, this dataset contained additional features about the participant context (\textbf{C$_{ps}$: }app usage, radius of gyration, screen/battery charging events). The dataset contained 3278 reports (1911 meals, 1367 snacks). The average age of study participants was 23.4 years, and the cohort had 44\% men and 56\% women. A more detailed feature summary with naming conventions is available in \cite{Meegahapola2020}.

\textbf{Data Pre-Processing:}\label{subsec:pre-processing}
During the feature extraction phase, standard datasets were created with one entry per eating event using a similar procedure to that given in \cite{Biel2018}. For both datasets, if the eating event occured at time T, we aggregated sensing data from T-$\alpha$ to T+$\alpha$ ($\alpha_{CH}$ = 2 hours as chosen in \cite{Biel2018}, $\alpha_{MX}$ = 30 minutes, we present results for this value after examining performance of the model for different $\alpha$).

\noindent \textbf{Activity (CH and MX Datasets):} For both datasets, (CH-Dataset: step counts and activity levels from fitbit; MX-Dataset: phone accelerometer), initially, the physical activity related features were calculated for 10-minute slots throughout the day. For the CH-Dataset: the total, median, mean, and standard deviation (\textit{sd}) values of these features were calculated for $\alpha_{CH}$ before (\textit{bef}) and after (\textit{aft}) each eating event using 10-minute based values. For MX-Dataset: features were derived from three axes of the accelerometer sensor using absolute (\textit{abs}) values and real values for $\alpha_{MX}$ before (\textit{bef}) and after (\textit{aft}) each eating event. Then, for the time window corresponding to the eating event, mean of feature values was calculated using 10-minute based values. 

\noindent \textbf{Apps (MX-Dataset):} We selected the ten most frequently used apps in the dataset. Then, during the eating time window, we determined whether each of the apps have been used or not, hence resulting in binary values for all app related features. 

\noindent \textbf{Location (CH and MX Datasets):} using location traces, we calculated radius of gyration \cite{Yue2014, Barlacchi2017} within the time period associated to the eating episode. 

\noindent \textbf{Screen (MX-Dataset):} using screen-on/off time slots, we calculated the screen-on time during the hour of consideration, and also the number of times the screen was turned on, similar to prior literature \cite{Abdullah2016}. 

\noindent \textbf{Battery (MX-Dataset):} we calculated the average battery level during the hour of consideration and also whether any charging events were detected during the time of eating episode. 

%


%% file: 03.methodology.tex
\section{Methodology and Results}\label{sec:methodology}

\subsection{Descriptive Analysis}\label{subsec:method:descriptive}

\begin{figure}
    \includegraphics[width=0.5\textwidth]{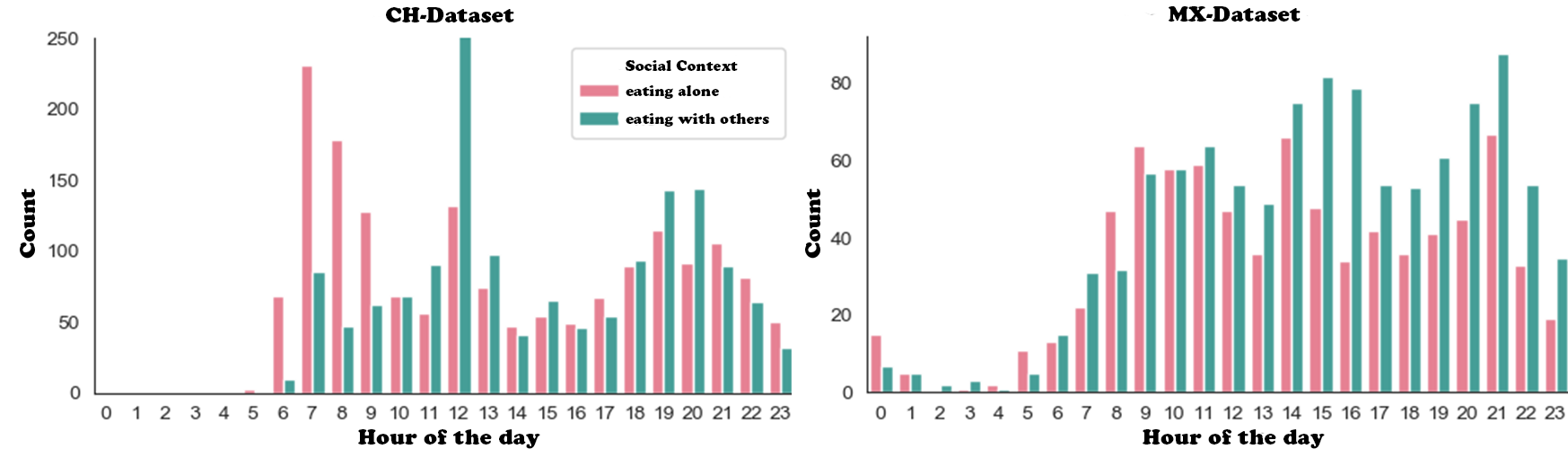}
    \caption{Temporal variation for social context of eating}
    \label{fig:time}

\end{figure}

\textbf{Temporal Variations.} Figure~\ref{fig:time} shows the temporal variation of the reported eating episodes for different social contexts of eating in both datasets. For both figures, three peaks can be seen that correspond to breakfast, lunch, and dinner, even though times at which these peaks occur are different in the two datasets. This could be due to cultural differences between the students in the two countries. Note that students in MX often live with their family while attending college, while this is less common with CH students. In the CH-Dataset, the time period from 6.00AM to 9.00AM (breakfast) results in significantly high number of {eating-alone} episodes compared to {eating-with-others} episodes. However, in the MX-Dataset, breakfast peak occurs later closer to 9.00AM to 11.00AM and the differences in terms of social context of eating are minimal, even though it still favors {eating-alone}. In CH-Dataset, a significantly high number of eating-with-others episodes are present during the lunch peak, and a similar pattern can be seen in the MX-Dataset as well. This could be because students were eating in the university, with their friends. In the CH-Dataset, dinner episodes are more or less even in terms of social context. However, the MX-Dataset show that even the dinner episodes are reported in highly social contexts, again partly explained for the living- with-parents situation. Hence, as a summary, Swiss students have reported high number of eating-alone episodes for breakfast, high number of eating-with-others episodes for lunch, and evenly distributed dinner episodes. For the MX students, except for the slight lean towards eating-alone in the morning, most other eating episodes have been reported to be with others. This could also be justified by prior research in psychology that has shown that Mexicans are highly social \cite{PamirezEsparza2009}.

\begin{figure}
    \centering
    \includegraphics[width=0.5\textwidth]{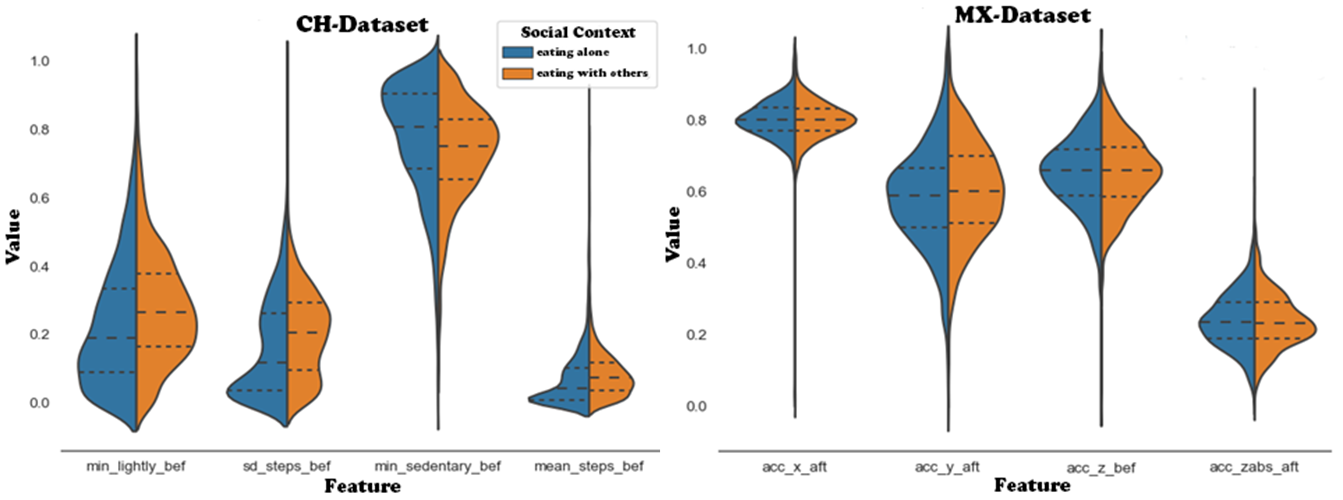}
    \caption{Violin plots for selected activity features}
    \label{fig:activity}

\end{figure}

\textbf{Passive Activity Variations.} Figure~\ref{fig:activity} shows differences in distributions of sensed activity levels for the two social contexts of eating, for few selected features (due to space limitations) using Violin plots \cite{violinplot2020}. In the CH-Dataset, activity level features captured using the wearable show significant mean differences for all four features shown here. For example, the distributions of feature \textit{mean steps bef} (see Section~\ref{sec:datasets} and \cite{Meegahapola2020} for feature descriptions) are different in terms of the shape, where eating-alone distribution is highly skewed towards lower values, meaning that activity levels are low for eating-alone episodes. A similar pattern can be seen for features such as \textit{min lightly bef} and \textit{sd steps bef}. The feature \textit{min sedentary bef}, that corresponds to time spent in sedentary state, show higher values for eating-alone episodes. As a summary, lower physical activity levels around eating episodes correspond to eating-alone in the CH-Dataset. This is consistent with prior work that has shown that low physical activity levels correspond to less social contexts \cite{Ward2016, Shelton2011}. However, in the MX-Dataset, the activity levels were sensed via the smartphone, and the mean differences are smaller for both social contexts of eating. Hence, for the MX case, it is less clear whether there are significant activity level differences depending on eating social context. However, it should be noted that the features generated regarding activity levels (based on availability) from the two datasets are different (fitbit for CH, phone accelerometer for MX), and this could have influenced the results.

\subsection{Statistical Analysis}\label{subsec:method:statistic}

Table~\ref{tab:statistics} shows statistics such as t-statistic \cite{Kim2015}, p-value \cite{Greenland2016}, and cohen's-d (effect size) \cite{Lakens2013} for all the features in both datasets for two groups: eating-alone and eating-with-others. The objective is to identify features that discriminate between the two social contexts. In the table, the features are ordered by the descending order of cohen's-d value. We calculated cohen's-d \cite{Rice2005} to help understand the statistical significance of the features because p-values are not sufficiently informative \cite{Yatani2016, Lee2016}. To interpret cohen's-d, we used a commonly used rule-of-thumb: small effect size = 0.2, medium effect size = 0.5 and large effect size = 0.8. Moreover, we calculated 95\% confidence interval for cohen's-d. A confidence interval that includes zero suggests statistical non-significance \cite{Lee2016}.

In the CH-Dataset, the feature with highest cohen's-d (medium effect size) is \textit{time since last meal}, which is derived from temporal aspects regarding food consumption. Moreover, physical activity features such as \textit{min lightly bef}, \textit{sd steps bef}, \textit{min sedentary bef}, and \textit{mean steps bef} that were derived from activity level by users before the eating event show cohen's-d values larger than 0.2. Moreover, all the features had confidence intervals that did not include zero. Other two features in the top ten are location and time, that also have cohen's-d values larger than 0.2. As a summary, the CH-Dataset contained several features  derived from fitbit that show discriminating signs between eating-alone and eating-with-others episodes.

When considering the MX-Dataset, the feature with highest cohen's-d (0.37) was \textit{location}. However, the only other feature from this dataset that had a cohen's-d higher than small effect size was \textit{time} (cohen's-d = 0.25). Even though several passive sensing features related to activity levels were in the top ten, only two features (\textit{acc y aft} and \textit{acc z abs aft}) had confidence intervals that did not include zero. Another passive interaction sensing modality that had a closer to small effect size was charging events with a cohen's-d of 0.16. Moreover, two app usage related passive sensing features (\textit{app google search} and \textit{app spotify}) were in the top ten. However, both these had cohen's-d confidence intervals including zero. As a summary, these results show that passive sensing features that quantify the activity levels, time of eating, and location have shown signs of discriminating capability between eating-alone and eating-with-others episodes in both datasets. 

\begin{table*}[t]
        \small
        \centering
        \caption{Comparative statistics of top 10 features across classes \textit{eating-alone} and \textit{eating-with-others}: t-statistic, p-value, and cohen's-d with 95\% confidence intervals. Features are sorted based on the decreasing order of cohen's-d; 95\% confidence interval of cohen's-d CI includes 0 = $^{x}$; when considering p-values, p<0.0001=*, p<0.001=**, p<0.01=***.}
               \begin{imageonly}
        \resizebox{0.85\textwidth}{!}{%
        \begin{tabular}{l l l l l l l l l }
        
        \hline 
        
        \rowcolor{Gray}
        \textbf{Feature} &
        \textbf{Feature Group} &
        \multicolumn{2}{c}{\textbf{CH-Dataset}} &
        &
        \textbf{Feature} &
        \textbf{Feature Group} &
        \multicolumn{2}{c}{\textbf{MX-Dataset}}
        \\
        
        \arrayrulecolor{Gray} \specialrule{6pt}{0pt}{-6pt} \arrayrulecolor{black}
        \cmidrule{3-4}
        \cmidrule{8-9}

        \rowcolor{Gray}
        & 
        &
        \textbf{cohen's-d} &
        \textbf{t-statistic} &
        & 
        &
        &
        \textbf{cohen's-d} &
        \textbf{t-statistic} 
        \\ \hline 

        time since last meal &
        T&
        0.57276 & 
        16.77194* & 
        &
        location &
        C$_{ps}$&
        0.37116 & 
        5.86325*
        \\
        
        \rowcolor{gray!5}
        min lightly bef &
        A$_{fb}$&
        0.32612 & 
        9.54787*& 
        &
        time &
        T&
        0.25114 & 
        4.02973*  
        \\
        
        sd steps bef &
        A$_{fb}$&
        0.31644 & 
        9.27218* & 
        &
        charging event &
        C$_{ps}$&
        0.16151 & 
        2.70769** 
        \\
        
        \rowcolor{gray!5}
        location &
        C$_{ps}$&
        0.26917 & 
        7.87712* & 
        &
        acc y aft &
        A$_{sp}$&
        0.15354 & 
        2.45258  
        \\
        
        min sed bef &
        A$_{fb}$&
        0.26841 & 
        7.86013* & 
        &
        acc y abs aft &
        A$_{sp}$&
        0.12842 & 
        2.05463 
        \\
        
        \rowcolor{gray!5}
        time &
        T&
        0.21020 & 
        6.15727* & 
        &
        app google search &
        C$_{ps}$&
        0.11903$^{x}$ & 
        1.87712 
        \\
        
        concurrent activity &
        C$_{sr}$&
        0.20756 & 
        6.09132* & 
        &
        acc z bef &
        A$_{sp}$&
        0.11629$^{x}$ & 
        1.86609  
        \\
        
        \rowcolor{gray!5}
        mean steps bef &
        A$_{fb}$&
        0.20101 & 
        5.88661* & 
        &
        acc z abs aft &
        A$_{sp}$&
        0.10989$^{x}$ & 
        1.76660 
        \\
        
        tot steps bef &
        A$_{fb}$&
        0.20079 & 
        5.88032* & 
        &
        screen on &
        C$_{ps}$&
        0.09536$^{x}$ & 
        1.51781 
        \\
        
        \rowcolor{gray!5}
        min lightly aft &
        A$_{fb}$&
        0.12267 & 
        3.59117*** & 
        &
        app spotify &
        C$_{ps}$&
        0.09453$^{x}$ & 
        1.48093 
        \\ \hline 
        
        \end{tabular}
        }
        \end{imageonly}
        \label{tab:statistics}
\end{table*}

\subsection{Inferring Eating Episodes: Alone or With Others}\label{subsec:method:inference}

The goal of the 2-class inference task was to use different subsets of features in the training set, and calculate the accuracy, precision, and recall. The target binary variable was eating-alone vs. eating-with-others, which indicates this fundamental aspect of eating. Moreover, we used random forest classifiers (RF) with ntree values between 100 - 150. We followed leave k-participants out strategy for all the experiments when preparing the dataset, where training and testing splits did not have data from the same user. Moreover, when preparing the dataset, we made sure that the classes are balanced by upsampling the minority classes to get balanced datasets (similar to \cite{Biel2018}). Further, we conducted the experiment for different feature groups, and feature group combinations such as (a) A: these are features generated using the activity data from fitbit wearables and smartphones. Features in this group are passively sensed, hence not needing any user interaction; (b) A+T: when temporal features such as time of the day and time since last food intake are combined with activity data, it provides a temporal variation of the activity levels; and (c) A+T+C$_{ps}$: this feature group contains only passive sensing features that are activity data and contextual data, and time of eating; (d) A+T+C$_{sr}$: this feature group contains only passive activity sensing features and contextual data that were self reported; (e) A+T+C$_{ps}$+C$_{sr}$: this combines all the available passive sensing and self-report features from wearables and smartphones to conduct the inference task. The baseline for experiments is 50\% since the classes were balanced in the inference task.

Results are summarized in Table~\ref{tab:inference_results}. As shown there, by only using activity data captured via wearables and smartphones, the models reached accuracies of 75.54\% and 70.57\% for the CH-Dataset and MX-Dataset, respectively. These accuracies are considerably increased when using temporal features. The A+T+C$_{ps}$ feature group shows how models that primarily use passive sensing data (even though T is self-reported, prior work has shown that the time of eating can be inferred to some extent with mobile sensing \cite{Thomaz2015v2, Bedri2017}. Hence, this feature group combination could be considered as near-passive.) without high user interaction to input details regarding details regarding food type or calorie levels, can infer eating social context with accuracies of 80.89\% and 77.73\% for CH-Dataset and MX-Dataset, respectively. For this feature group, features such as time, time since last meal, location, and other activity related features were among the top five for both datasets, when considering feature importance values derived from the RFs. Moreover, when all the features used in mobile food diaries with sensing capabilities are considered, the inference accuracies reached higher values of 90.88\% and 84.08\% for the CH and MX datasets, respectively. These results show that precise wearable sensing and even smartphone sensing (that can be obtained regardless of the smartphone type or brand) are both useful to infer eating-alone vs. eating-with-others episodes. This could be seen as a first step towards enabling holistic mobile food diaries with reduced user burden, by inferring attributes that are typically captured with self-reports.

\begin{table}[t]
        \small
        \centering
        \caption{\textbf{\textit{Eating-alone} vs. \textit{eating-with-others} inference task}}
        \begin{imageonly}
        \resizebox{0.4\textwidth}{!}{%
        \begin{tabular}{l l l l l}
\hline 
        
        \rowcolor{Gray}
          &
        \textbf{Feature Group} &
        \textbf{Accuracy} &
        \textbf{Precision} &
        \textbf{Recall} 
        \\ \hline 
        
        &
        Baseline &
        50.00\% & 
        - & 
        -  
        \\ \hline

        \multirow{5}{*}{\rotatebox[origin=c]{90}{\footnotesize{CH-Dataset}}}&
        A$_{fb}$ &
        75.54\% & 
        75.52\% & 
        75.53\%  
        \\
        
        &
        A$_{fb}$+T  &
        80.31\% & 
        80.35\% & 
        80.29\%  
        \\

        &
        \CC{15}A$_{fb}$+T+C$_{ps}$  &
        \CC{15}80.89\% & 
        \CC{15}80.96\% & 
        \CC{15}80.90\% 
        \\
        
        &
        A$_{fb}$+T+C$_{sr}$  &
        89.97\% & 
        90.31\% & 
        89.56\%  
        \\
        
        &
        A$_{fb}$+T+C$_{ps}$+C$_{sr}$  &
        90.88\% & 
        91.11\% & 
        90.69\%  
        \\ \hline 

        \multirow{5}{*}{\rotatebox[origin=c]{90}{\footnotesize{MX-Dataset}}}&
        A$_{sp}$  &
        70.57\% & 
        71.11\% & 
        71.10\%  
        \\
        
        &
        A$_{sp}$+T  &
        72.30\% & 
        73.03\% & 
        72.92\%  
        \\

        &
        \CC{15}A$_{sp}$+T+C$_{ps}$  &
        \CC{15}77.73\% & 
        \CC{15}77.74\% & 
        \CC{15}77.77\%  
        \\
        
        &
        A$_{sp}$+T+C$_{sr}$   &
        78.29\% & 
        78.59\% & 
        78.31\%  
        \\
        
        &
        A$_{sp}$+T+C$_{ps}$+C$_{sr}$   &
        84.08\% & 
        84.24\% & 
        84.03\%  
        \\ 
        \hline 
        
        \end{tabular}
        }
                \end{imageonly}
        \label{tab:inference_results}
\end{table}

%% file: 05.conclusion.tex
\section{Conclusion}\label{sec:conclusion}

In this paper, we investigated whether mobile sensing features could help identify the social context of eating episodes, using datasets from two  countries (Switzerland and Mexico). We identified dataset features that help discriminate between two sociability levels ({eating-alone} and {eating-with-others}), and finally, we tackled a novel ubicomp task to infer eating-alone vs. eating-with-others with accuracies of the range 77\% - 81\% by using passive sensing data from wearables and smartphones. Even when only activity related data sensed via wearables and smartphones are considered, binary sociability can be inferred with accuracies of 70.57\% and 75.54\% for the MX-Dataset and CH-Dataset, respectively. These accuracies can be increased to the range of 84\%-91\% by combining other self-reported attributes. These results show initial progress towards inferring traditionally self-reported attributes such as social context of eating, which is useful to design mobile food diaries that reduce user burden and provide context-aware functionalities. This analysis could also be a first step towards understanding more complex social contexts (with friends or with family), which remain open for future investigation. Finally, it would also be interesting to examine how approaches such as transfer learning can be used to examine attributes from datasets of similar nature (e.g. about food consumption behavior), but have certain differences (e.g. passive sensing data captured from Switzerland and Mexico using slightly different approaches).  

\balance


\begin{acks}
This work was funded by the European Union’s Horizon 2020 WeNet project, under grant agreement 823783. D. Gatica-Perez acknowledges the former support of EPFL Integrative Food and Nutrition Center, through the Understanding Eating Routines in Context project.
\end{acks}